\documentstyle[12pt,aasms4]{article}

\def\Msun{\mbox{M$_{\odot}$}}

\def\BmV0{\mbox{(B-V)$^{\rm o}$}}
\def\VmK0{\mbox{(V-K)$^{\rm o}$}}
\def\MV0{\mbox{M$_{\rm V}^{\rm o}$}}

\def\Msun{\mbox{M$_{\odot}$}}

\def\etal{\mbox{{\it et al.}}}

\def\third{{3$^{\rm rd}$}}
\def\gtaprx{ \mathrel{  \vcenter{
                        \offinterlineskip \hbox{$>$}
                        \kern 0.3ex \hbox{$\sim$}    } } }
\def\ltaprx{ \mathrel{  \vcenter{
                        \offinterlineskip \hbox{$<$}
                        \kern 0.3ex \hbox{$\sim$}    } } }
\lefthead{}
\righthead{}
\begin{document}

\title{ 
Probing the Neutron-Capture Nucleosynthesis History of Galactic Matter}

\author{
James W. Truran\altaffilmark{1},
John J. Cowan\altaffilmark{2}, 
Catherine A. Pilachowski\altaffilmark{3}
and Christopher Sneden\altaffilmark{4}} 

\altaffiltext{1}{Department of Astronomy and Astrophysics, Enrico Fermi 
Institute, University of Chicago, 933 E. 56th Street, Chicago, IL 60637;
truran@nova.uchicago.edu} 
\altaffiltext{2}{Department of Physics and Astronomy,
University of Oklahoma, Norman, OK 73019; 
cowan@mail.nhn.ou.edu}
\altaffiltext{3}{Astronomy Department, Indiana University Bloomington,
727 E. 3$^{rd}$ St., Bloomington, IN 47405; catyp@astro.indiana.edu}
\altaffiltext{4}{Department of Astronomy and McDonald
 Observatory, University of Texas, Austin, TX 78712;
chris@verdi.as.utexas.edu}

\begin{abstract}

     The heavy elements formed by neutron capture processes have an interesting
history from which we can extract useful clues to and constraints
upon both the characteristics of the processes themselves and  
the star formation and nucleosynthesis history of Galactic matter.
Of particular interest in this regard 
are the heavy element compositions of extremely metal-deficient stars. 
At metallicities [Fe/H]
 $\leq$ --2.5, the elements in the mass region past
barium (A $\gtaprx$ 130-140) have been found (in non carbon-rich stars) 
to be pure $r$-process
products. The identification of an environment provided by massive stars and
associated Type II supernovae as an $r$-process site seems compelling.
Increasing levels of heavy $s$-process (e.g., barium) 
enrichment with increasing
metallicity, evident in the abundances of more metal-rich halo stars and
disk stars,  reflect the delayed contributions from the low- and
intermediate-mass 
(M~$\sim$ 1-3 M$_\odot$) stars that provide 
the site for the main 
$s$-process nucleosynthesis component during the AGB phase
of their evolution. New abundance data in the mass 
region 60 $\ltaprx$ A $\ltaprx$ 130 is providing insight into the 
identity of possible alternative $r$-process sites. 
We review recent observational studies of
heavy element abundances both in low metallicity halo stars and in disk stars, 
discuss the observed trends in light of nucleosynthesis theory, 
and explore some implications of these results for  
Galactic chemical evolution, nucleosynthesis, and nucleocosmochronology. 

\end{abstract}


\section{INTRODUCTION}

Element abundance patterns in metal-poor halo field stars and globular
cluster stars play a crucial role in guiding and constraining theoretical
models of Galactic nucleosynthesis.
These patterns can also provide significant clues to the natures of the 
nucleosynthesis mechanisms themselves. 
Nowhere is this more true than for the case of the neutron-capture 
($n$-capture) processes
that are understood to be responsible for the synthesis of the bulk of the
heavy elements in the mass region A $\gtaprx$ 60: the 
$s$-process and the
$r$-process. Nucleosynthesis theory identifies quite different 
physical conditions and astrophysical
sites for these two distinct processes. 
$r$-Process nuclei are effectively 
{\it primary} nucleosynthesis products, formed 
under dynamic conditions in an environment associated
with the evolution of massive stars (M $\gtaprx$ 10 M$_\odot$) to supernova
explosions of Type II and the formation of neutron star remnants.
$s$-Process nuclei are understood to be products of neutron captures
on preexisting silicon-iron ``seed'' nuclei, occurring under hydrostatic 
burning conditions both in the helium
burning cores of massive stars and particularly in the thermally pulsing
helium shells of asymptotic giant branch (AGB) stars. In this picture, the
first heavy (A $\gtaprx$ 60) elements introduced into the interstellar gas
component of our Galaxy are expected to have been $r$-process 
nuclei formed
in association with massive stars, on time scales $\tau_{star}$ 
$\ltaprx$ 10$^8$ 
years. Most of the $s$-process nuclei, 
on the other hand, are first introduced into 
the ISM on time scales ($\sim$ 10$^9$ years) characteristic of the lifetimes 
of their stellar progenitors (M $\sim$ 1-3 M$_\odot$).

That the general features of this simple model are correct is confirmed by
the finding that $r$-process contributions 
dominate the heavy element abundances
in extremely metal-poor halo and globular cluster 
stars, while significant $s$-process contributions are first 
identifiable at metallicities of order [Fe/H]
\footnote{
We adopt the usual spectroscopic notations that
[A/B]~$\equiv$ log$_{\rm 10}$(N$_{\rm A}$/N$_{\rm B}$)$_{\rm star}$~--~log$_{\rm 10}$(N$_{\rm A}$/N$_{\rm B}$)$_{\odot}$, and
that log~$\epsilon$(A)~$\equiv$ log$_{\rm 10}$(N$_{\rm A}$/N$_{\rm H}$)~+~12.0,
for elements A and B.
Also, metallicity will be assumed here to be equivalent to the stellar
[Fe/H] value.} 
$\approx$ --2.  
Observational constraints on $r$-process nucleosynthesis 
sites are  examined in \S 3. The implications of the observed scatter in 
the level of $r$-process nuclei 
relative to iron and trends in the heavy 
element abundances as a function of [Fe/H] over the early star formation 
history of the Galaxy are considered in \S 4. 
We also consider the use of the abundance 
data - particularly that involving the nuclear chronometers $^{232}$Th 
and $^{238}$U - for cosmochronology (\S 5). 
Finally, a brief examination of neutron-capture nucleosynthesis at 
low metallicities is presented in \S 6. We note there the importance
of the carbon-rich stars. Although  this review does not include a 
detailed examination of this class of metal-poor stars,
the abundance patterns 
in these stars may provide insight into the earliest phases of 
Galactic nucleosynthesis. 
First, in order to provide a basis for our subsequent discussions of the
heavy element abundance patterns in the stellar populations of our Galaxy,
we briefly review in the next section the current status of theoretical models 
for $s$-process and $r$-process nucleosynthesis.
Throughout this paper, our emphasis will be on (first) providing the best 
observational data currently available concerning heavy element abundances 
in stars in our Galaxy and (second) 
the use of this data to educate us concerning 
the physical characteristics, sites, and time scales of $s$-process 
and $r$-process nucleosynthesis.

\section{S-PROCESS AND R-PROCESS SITES AND MECHANISMS }

We are concerned in this review with the interpretation of the
heavy element (A$\gtaprx$60) abundance patterns observed in diverse stellar
populations in the context of nucleosynthesis theory. 
Following upon the early discussions of nucleosynthesis mechanisms
by Cameron (1957) and Burbidge {\it et al.} (1957), we understand that most
of the heavy elements are formed in two processes involving neutron captures:
the $s$-process and the $r$-process. These two broad divisions 
are distinguished on the basis of relative lifetimes for neutron captures
($\tau_n$) and electron decays ($\tau_\beta$). The condition that
$\tau_n$$>$$\tau_\beta$, where $\tau_\beta$ is a characteristic lifetime for
beta-unstable nuclei near the valley of beta stability, ensures that as
captures proceed the $n$-capture path will itself remain close to the
valley of beta stability. This defines the 
$s$-process. In contrast, when
$\tau_n$$<$$\tau_\beta$, it follows that successive neutron captures will
proceed into the neutron-rich regions off the beta-stable valley. Following
the exhaustion of the neutron flux, the capture products approach the position
of the valley of beta stability by beta decay, forming the 
$r$-process nuclei.
Using experimental determinations of the neutron-capture cross sections and
the smooth behavior  of the product of the abundance and cross section
($\sigma_{n,\gamma}$N$_s$) for nuclei lying along the 
$s$-process path,
K\"appeler {\it et al.}(1989) have identified and extracted (Figure 1) 
the $s$-process and $r$-process 
patterns characterizing solar system matter. 

The significant point, for our
purposes, is that these patterns are readily distinguishable. 
This immediately implies the following: 

\begin{itemize}

\item{} If we can identify stellar environments in which either the 
$s$-process 
or the $r$-process contributions dominate, 
we can use this information to 
constrain the detailed characteristics of the corresponding nucleosynthesis 
mechanism.

\item{} If we can distinguish $s$-process and 
$r$-process elemental contributions, 
we can use stellar abundance data to trace the chemical evolution of such 
processes over Galactic history.  

\end{itemize}

In fact, we have been successful in both quests. Patterns of $s$-process
elements are observed to be enriched in some red giant stars, reflecting 
{\it in situ} neutron capture nucleosynthesis. Indeed, it was the 
identification of technetium in the atmospheres of red giants by 
Merrill (1952) that provided a strong early clue to the fact that heavy 
element synthesis occurs in stars. Similarly, as we will see, the 
$r$-process 
distribution characteristic of solar system matter is unambiguously  
identifiable in the spectra of extremely metal-poor stars. The complicated 
chemical evolutionary histories of the two neutron capture processes 
can then be traced by examining, for example, the history of the 
barium (formed predominantly in the $s$-process) to europium (
nearly an r-only 
element) ratio as a function of [Fe/H]. We will return to this issue in \S 4.

\subsection{s-Process Nucleosynthesis}

Theory has proposed the existence of several quite different astrophysical sites
for the 
operation of the two neutron-capture nucleosynthesis mechanisms. In fact, 
as we shall 
see, it would now appear that there may be (at least?) 
two identifiable and distinct 
components (environments) for both $s$-process and 
$r$-process nucleosynthesis. 
For the case of the $s$-process, the two astrophysical environments are:

\begin{itemize}

\item{} \underbar{The helium burning cores of massive stars} 
(M $\gtaprx$ 10 M$_\odot$) 
provide an environment in which the $^{22}$Ne($\alpha$,n)$^{25}$Mg reaction 
can operate to produce 
$s$-process nuclei through the mass region A $\approx$ 90
(the {\it ``weak''} component). First studied by Peters (1968) and by Lamb 
{\it et al.} (1977), this process can in principle provide a source of the 
lightest $s$-process nuclei during the 
early stages of Galactic evolution (as 
soon as significant production of iron seed nuclei has occurred). Recent  
studies (Raiteri, Gallino, \& Busso 1992; Baraffe, El Eid, \& Prantzos 1992;
Heger et al. 2002) reveal that the efficiency of production of 
$s$-process nuclei decreases 
at low metallicities (below [Fe/H] $\approx$ -2) due to the increased 
competition arising from the elevated levels of abundance of nuclei from 
Ne to Ca relative to iron. We will return to this issue when we discuss the 
evolution of $s$-process abundances in \S 4. 

\item{} \underbar{The thermally pulsing helium shells of asymptotic giant 
branch stars}  
provide an environment in which the $^{13}$C($\alpha$,n)$^{16}$O reaction can
operate to produce 
$s$-process nuclei in the heavy region through to lead and
bismuth (the {\it ``main''} component). First identified as a promising 
$s$-process 
site by Schwarzschild and H\"arm (1967), this site has since been 
investigated extensively by a number of authors (see, e.g., the recent 
review by Busso, Gallino, \& Wasserburg (1999). This 
$s$-process site (the 
{\it ``main''} $s$-process component), which 
is understood to be responsible for the synthesis of the  
$s$-process 
nuclei (A $\gtaprx$ 90), operates predominantly in low mass
(M $\sim$ 1-3 M$_\odot$) stars.  
The lifetimes of stars in this mass range are 
typically ($\tau$ $\gtaprx$ 10$^9$ years), significantly greater than the 
lifetimes of the massive stars (M $\gtaprx$ 10 M$_\odot$) that are the site 
of formation of both the weak $s$-process component and the 
$r$-process heavy nuclei. These differential time scales for the 
return of the products of $s$-process and $r$-process 
nucleosynthesis to the interstellar medium (ISM) imply a rather complicated 
chemical evolutionary history for the elements in the mass range 
A $\gtaprx$ 60--70. We will explore this history in greater detail in 
section 4. 

\end{itemize}

\subsection{r-Process Nucleosynthesis} 

The site(s) of the astrophysical 
$r$-process(es) remains a challenge to theorists (see the earlier 
review by Cowan, Thielemann \& Truran 1991). 
While the general nature of the 
$r$-process and of its contributions to the
abundances of heavy elements in the mass range through uranium and
thorium are generally understood, the details remain to be worked out. 
There are considerable uncertainties associated both with the basic nuclear
physics of the $r$-process - which involves the 
$n$-capture, $\beta$-decay,
and fission properties of unstable nuclear species far from the region of
$\beta$-stability - and with the characteristics of the stellar or supernova
environments in which $r$-process synthesis occurs. 
Until recently, it was at
least reassuring that, in contrast to $s$-process, it appeared a single 
$r$-process
site was involved. Observations reviewed in  section 3 now suggest that
this is not true - rather, there must be distinct classes of 
$r$-process events
operating in the mass regimes above and below masses A $\approx$ 130-140. 

Promising and studied sites of $r$-process 
nucleosynthesis include the following: 

\begin{itemize}  

\item{} The $r$-process model that has 
received the greatest study in recent 
years involves a high-entropy (neutrino-driven) wind from a Type II 
supernova (Woosley {\it et al.} 1994; Takahashi, Witti, \& Janka 1994). The 
attractive features of this model include the facts that it may be a natural 
consequence of the neutrino emission that must accompany core collapse in 
Type II events, that it would appear to be quite robust, and that it is 
indeed associated with massive stars of short lifetime. Recent calculations 
have, however, called attention to a critical problem associated with this 
mechanism: the entropy values predicted by current Type II supernova models 
are too low to yield the correct levels of production of both the lighter 
and heavier $r$-process nuclei. 

\item{} The conditions estimated to characterize the decompressed ejecta from 
neutron star mergers (Lattimer {\it et al.} 1977; Rosswog et al. 1999) 
may also be compatible with the production of an 
$r$-process abundance pattern generally 
consistent with solar system matter. 
The most recent numerical study of $r$-process 
nucleosynthesis in matter ejected 
in such mergers (Freiburghaus, Rosswog, \& Thielemann 1999) show specifically 
that the $r$-process heavy nuclei in the mass range A $\gtaprx$ 130-140 and 
produced in solar proportions. 
Here again, the association with a massive star/Type II supernova environment 
is consistent with the early appearance of 
$r$-process nuclei in the Galaxy and 
the mechanism seems quite robust. A potential problem with this source, 
pointed out by Qian (2000), is that the observed frequency of such events in 
the Galaxy, lower by a factor $\approx$ 100 than the frequency of Type II 
supernovae, demands that a high mass of 
$r$-process matter ejected per event. 
He argues that this is inconsistent with the level of scatter of 
[$r$-process/Fe] observed in halo stars.

\item{} LeBlanc \& Wilson (1970) have considered possible ejection of 
neutronized material in magnetized jets from collapsing stellar cores. 
This mechanism, which has most recently been examined by Cameron (2001), 
is again tied to massive star/Type II SNe environments. 

\item{} The helium and carbon shells of massive stars undergoing supernova 
explosions can also give rise to significant neutron production, via such 
reactions as $^{13}$C($\alpha$,n)$^{16}$O, $^{18}$O($\alpha$,n)$^{21}$Ne, 
and $^{22}$Ne($\alpha$,n)$^{25}$Mg, involving residues of hydrostatic 
burning phases (Truran, Cowan, \& Cameron 1978; Thielemann, Arnould, \& 
Hillebrandt 1979; Blake {\it et al.} 1981). Here again, the site is associated 
with a Type II supernova event. The weakness of this environment, as revealed 
by the calculations cited above, is the fact that the expected neutron exposure 
is not consistent with the production of the entire range of 
$r$-process nuclei 
through uranium and thorium. Recent calculations (Truran \& Cowan 2000)  
suggest that these supernova conditions may be consistent 
with the production of the lighter $r$-process nuclei,
 through the mass range 
A $\approx$ 130-140. As we will discuss in the next two sections, this 
might explain the lighter $n$-capture 
 element abundance patterns observed in certain 
low-metallicity stars. However, preliminary calculations by 
Heger {\it et al.} (2002) indicate, rather, that the available neutron flux 
for these environments may not be sufficient to synthesize even these lighter
$r$-process elements. The question regarding the 
contributions of the helium and carbon shells of massive stars to solar 
system and galactic $r$-process abundances thus remains open.

\end{itemize} 

In summary, the theoretical picture indicates the inherent complexity of 
the heavy element nucleosynthesis history. 
There is increasingly strong observational evidence that the 
$r$-process
isotopes identified in solar system matter are in fact the products of two
distinct classes of $r$-process events - for the regions 
A $\gtaprx$ 130-140 and A $\ltaprx$ 130-140. Supernovae of a certain mass range,
or neutron star mergers, 
appear to be 
the most promising candidates for production of nuclei in the mass regime 
A $\gtaprx$ 130-140. 
Shock 
processing of the helium and/or carbon shells in Type II supernovae may 
produce the $r$-process nuclei in the mass range A $\ltaprx$ 130-140. 
The 
(seemingly) lower level of the $r$-process abundances  for  A $\ltaprx$ 130-140 
in extremely 
metal deficient halo stars like CS 22892-052 (Sneden {\it et al.} 2000a) may 
simply be attributable to the operation of the helium/carbon shell 
mechanism at low metallicities. 
Alternatively, perhaps supernovae of different mass range 
or frequency than those
that produced the heavier $r$-process elements might be responsible
for the synthesis of nuclei with A 
$\ltaprx$ 130-140.
There have also been suggestions (Cameron 2001) 
that the entire abundance distribution
could be synthesized in a certain type of core-collapse 
supernova.
Note that 
in all cases the site is associated with massive stars and can therefore 
enrich the interstellar medium on a relatively rapid time scales.

\section{EVIDENCE FOR R-PROCESS NUCLEOSYNTHESIS AT EARLY GALACTIC EPOCHS}

Studies of element abundances in the oldest and most metal-deficient stars 
in our Galaxy are extremely important because 
they serve as tests of nucleosynthesis 
theories and of models of Galactic chemical evolution, and can provide critical 
inferences concerning the star formation histories of stellar populations. 
Reviews of overall
 abundance patterns as a function of metallicity [Fe/H] have been 
provided by Spite \& Spite (1985), Wheeler, Sneden, and Truran (1989), and 
McWilliam (1997). In this review, we concentrate specifically on the heavy 
element products of $s$-process and $r$-process 
nucleosynthesis.

The fact that there exist real and systematic depletions in the 
$s$-process elements relative to iron in stars of low [Fe/H] was first 
emphasized by Pagel (1968). Observed trends in the abundances of the designated 
$s$-process nuclei were scrutinized by Tinsley (1979; see also 
Truran 1980) and found to be inconsistent with ``conventional theories.''
The clue to what was happening here was provided by the observations of Spite 
\& Spite (1978) of the europium abundances in metal poor stars. Europium is an 
element with two stable isotopes, the abundances of both of which are 
dominated by their $r$-process contributions. The Spite \&  Spite 
data revealed that the Eu/Fe ratio was essentially solar (or higher), 
even for stars of very low metallicity (--3 $\ltaprx$ [Fe/H] $\ltaprx$ --2), 
leaving no doubt that $r$-process nucleosynthesis had occurred. Many 
of the ``anomalous'' trends in ``$s$-process'' abundances at low 
metallicities then became interpretable rather on the basis of their levels of 
production in the $r$-process (Truran 1981). 

This $r$-process interpretation has since been unambiguously 
confirmed by observations, beginning with studies of $n$-capture elements
in HD~122563 ([Fe/H] $\approx$ --2.7) and HD 110184 ([Fe/H] $\approx$ --2.5)
by 
Sneden \& Parthasarathy (1983), Sneden \&  
Pilachowski (1985) and with the larger survey by 
Gilroy {\it et al.} (1988).  
These papers all demonstrated  that 
the heavy element abundance patterns in very low-metallicity giants 
were consistent 
with solar system 
$r$-process, but not  
$s$-process, abundances.

Recent observational studies have served both to confirm and to make
clearer the presence of 
such patterns in low metallicity stars and to emphasize the extraordinary 
agreement of the observed $r$-process patterns with that of solar 
system matter. Both ground based (Sneden {\it et al.} 1996; Sneden {\it et al.} 
2000a; Burris {\it et al.} 2000; Westin 2000) and space based (Cowan 
{\it et al.} 1996; Sneden {\it et al.} 1998) observational studies have now 
confirmed that 
the abundances of the heavy $n$-capture products (A $\gtaprx$ 130-140)
in the most metal deficient halo field stars and globular cluster stars
([Fe/H] $\ltaprx$ -2.5) were formed in an  $r$-process event. This is
reflected in the abundance pattern for the ultra metal poor
([Fe/H] = -3.1), but $r$-process enriched ([$r$-process/Fe] $\approx$ 30-50),
halo star CS 22892-052 shown in Figure 2. Note the
truly remarkable agreement of the elemental abundance pattern in the region
from barium through at least iridium, and probably through lead
 with the solar system elemental pattern.

This comprehensive  level of agreement of the
metal-poor star heavy element abundance pattern with (solar system) $r$-process
abundances is seen in most of the  stars of [Fe/H] $\ltaprx$ --2.5 
with published detailed $n$-capture abundance patterns.\footnote{We note, 
however,  that  
the   observed stars tend to be $r$-process ``rich,'' and we do not have as 
much data for the $r$-process ``poor'' stars from [Fe/H] --2.5 to --3.0,
nor for metal-poor stars with metallicities below --3.0 (see discussion 
below in \S 6).}   
The
heavy element abundance patterns for the three $r$-process rich stars
CS 22892-052 (Sneden {\it et al.} 2000a), HD 115444 (Westin {\it et al.} 2000), 
and BD +17$^o$3248 (Cowan {\it et al.} 2002) are compared
with the solar system $r$-process abundances in Figure 3. The robustness of the 
$r$-process mechanism operating in the early Galaxy is reflected in the
spectacular agreement of these three stellar patterns (which represent the
nucleosynthesis products of, at most, a relatively small number of
earlier stars) with the solar
system $r$-process pattern (which represents the accumulated production
of $r$-process elements over billions of years of Galactic evolution).
Note the extraordinary agreement with solar system $r$-process abundances
over this range, for the eighteen elements for which abundance data is
available. These data provide conclusive evidence for the operation of
an $r$-process at the earliest Galactic epochs that synthesizes the heavy
$r$-process nuclei (barium and beyond), including the long lived isotopes
$^{232}$Th and $^{238}$U critical to dating. 
The identification with massive stars seems
compelling, although it is possible that neutron star mergers rather than
a supernova environment may be responsible.

A critical question arises here involving 
the abundances of $r$-process nuclei
less massive than barium. The incompatibility of the abundances of
the short lived radioactivities $^{107}$Pd and $^{129}$I in early solar
system matter with a model of ``uniform production'' that works well for
both the actinide chronometers and the short lived isotope $^{182}$Hf
led Wasserburg {\it et al.} (1996) to argue for a second
$r$-process component for the mass region 
below A $\approx$ 130-140. Studies of
abundances in the mass range from the iron region through barium in
low metallicity stars can be utilized to
provide significant input in this regard.
Returning to Figure 2  
we note that 
while there is remarkable agreement with solar $r$-process abundances for the 
the heavier nuclei, the abundance pattern in the mass regime below 
A $\approx$ 130-140 does not exhibit this consistency. This is evident as well 
for the halo star BD +17$^o$3248 (Cowan {\it et al.} 2002), as shown in 
Figure 4. 

Recent observational studies of the heavy element abundance patterns in 
globular clusters provide further evidence of the early dominance of the 
$r$-process contributions in Galactic matter. Here, we are particularly 
interested in the question as to whether the abundance history of 
[$r$-process/Fe] 
in globular clusters parallels that of the field halo stars. 
To date, there have been only a few detailed abundance studies of the 
heavy neutron capture elements in globular clusters 
(see, e.g. Ivans {\it et al.} 1999; 2001). One such study by
Sneden et al. (2000b) (see Figure 5) does indicate a pattern similar
to that seen in the field halo stars, and again consistent with the
scaled solar system $r$-process abundances. We also note the detection 
of the long-lived radioactive element thorium in several of the 
giants in M15. These detections hold implications for radioactive 
age determinations and cosmochronology, as discussed in section 5.  


\section{CHEMICAL EVOLUTION OF r-PROCESS and s-PROCESS ABUNDANCES}

In this section we examine more closely the abundance histories of individual
neutron capture elements as a function of metallicity. As we shall see, they 
serve to provide important information concerning the earliest stages of 
star formation and nucleosynthesis enrichment in the early Galaxy.

\subsection{Scatter in the Early Galaxy}

We first explore the  abundance scatter in the Galaxy. 
Early work by Gilroy {\it et al.} (1988) 
first proposed  that the stellar abundances of
$r$-process elements with respect to iron, particularly Eu/Fe, showed
a large scatter at low metallicities. Their work suggested 
that this  scatter appeared to diminish
with increasing metallicity. A more extensive study by
Burris {\it et al.} (2000) 
confirmed the very large star-to-star  scatter in the early Galaxy,
while studies of
stars at higher metallicities, involving mostly disk stars 
(Edvardsson {\it et al.} 1993; Woolf {\it et al.} 1995), show little scatter.  
In Figure 6, we plot the data from a number of these surveys
(Burris {\it et al.} 2000; Edvardsson {\it et al.} 1993; Woolf {\it et al.} 
1995; Jehin {\it et al.} 1999; McWilliam et al. 1995, McWilliam 1998). 
These studies, which cover a broad metallicity range of 
--3.5$\le$[Fe/H]$\le$+0.5,
include large numbers of stars and  hyper fine splitting in the
derivation of Ba abundances.

We have also plotted in Figure 6 
abundance determinations from several individual  stars with 
large overabundances of $r$-process elements, which we refer to as
$r$-process ``rich'' stars. These include  CS 22892-052 (Sneden et al. 2000a), 
HD 115444 (Westin et al. 2000), CS 31082-001 (Hill et al. 2002) and 
BD +17$^o$3248 (Cowan et al. 2002). 
Thus,  in some of those stars we see 
the ratio of the $r$-process element Eu, to Fe,  
with values 
reaching as high as [Eu/Fe] $\approx$ 50.
(Even though the total Eu abundance
is still less,  the ratio of this element to  
iron is much higher in some of these stars than in the Sun.) 
At a metallicity of [Fe/H]=--3 in Figure 6,
the abundance ratio of [Eu/Fe] varies by
more than two orders of magnitude. However, near [Fe/H] = --1 
the variation in [Eu/Fe] is reduced to less than 
a factor of $\sim$  five.
The star-to-star scatter illustrated in this figure 
can most easily be explained as due to local inhomogeneities 
resulting from contributions 
from individual nucleosynthetic events (e.g., supernovae), and strongly  
suggests an early, unmixed, chemically inhomogeneous Galaxy. 

$r$-Process abundances in low metal stars 
also provide further clues to the natures of 
the $r$-process  and $s$-process sites.
For example, 
the increasing level of scatter of [Eu/Fe]  with decreasing metallicity, 
most pronounced at values below [Fe/H] $\sim$ --2.0, makes clear that not
all early stars are sites for the formation of both 
$r$-process nuclei and
iron. In the absence of other sources for either iron peak nuclei or
$r$-process nuclei in the early Galaxy, 
this scatter is consistent with the
view that only a small fraction 
(2\%--10\%)  of the massive stars that produce iron
also yield $r$-process elements. 
Discussions of this issue may  be found in the papers by Fields, Truran, 
\& Cowan (2002) and Qian \& Wasserburg (2001). 

One other important trend is notable in Figure 6.
At higher metallicities, particularly for [Fe/H] $\simeq$ --1,
the values of [Eu/Fe] tend downward.
This reflects the effect of increasing iron production,
presumably from Type Ia supernovae, at higher Galactic metallicities. 
At very low metallicities high mass (and
rapidly evolving) Type II supernovae contribute to Galactic iron production.
The onset of the bulk of iron production from Type Ia supernovae
(with longer evolutionary time scales due to lower
mass progenitors) occurs only at higher [Fe/H]  and later Galactic times.


\subsection{Abundance Trends}

Studies of the time history of the heavy (neutron capture) elements
have historically been driven by observations. Early trends in the
$s$-process element patterns as a function of [Fe/H] by Pagel (1968) 
motivated theoretical consideration of such trends in the context of
models of Galactic chemical evolution (Truran and Cameron 1971; Tinsley 1980).
The great increase in abundance data for metal-poor stars over the past
decade has similarly motivated considerable theoretical activity.
The ratio [Ba/Eu], which reflects the ratio of $s$-process to
$r$-process elemental abundances, 
is displayed, in Figure 7,  as a function of
[Fe/H], for a large sample of halo and disk stars. Note that
at the lowest metallicities the [Ba/Eu] ratio clusters around the  
pure $r$-process value. 
The question then is what is the 
Galactic metallicity at which major contributions from $s$-process
nucleosynthesis began to generally  appear in the halo ISM?
Most $s$-process synthesis is associated with the AGB phases of
low  
(M $\sim$ 1-3 M$_\odot$) and intermediate-mass (M $\sim$ 3.5-8 M$_\odot$)
stars, 
whose evolutionary time scales are $\ge$ $\sim$10$^8$~yr.
Metallicity regimes with little or no detectable $s$-process contributions
are those resulting from the first waves of Galactic nucleosynthesis
that happened on faster time scales.
The approximate [Fe/H] indicating a general rise in the $s$-process
should tell us how much the first nucleosynthesis burst contributed
to overall Galactic metallicity.
With earlier  data, Gilroy et~al. (1988) suggested that the onset of major
Galactic $s$-process occurred at [Fe/H]~$\sim$~--2.3. 
Burris et~al. (2000) found evidence for $s$-processing in some stars
with  metallicities
as low as  [Fe/H]~$\sim$~--2.8, with most stars with
 [Fe/H]~$\sim$ $>$ --2.3 showing evidence for $s$-process 
contributions to the total $n$-capture abundances.
This fairly wide range of metallicity might indicate the various
contributions from both low-mass and intermediate-mass AGB stars 
with different evolutionary time scales for injecting $s$-process material
into the ISM.
These chemical evolution trends have also motivated extensive work on the 
problem of the early chemical evolution of the Galaxy 
(Mathews, Bazan, \& Cowan   1992;  
Ishimaru \& Wanajo 1999; Travaglio {\it et al.} 1999).


Another observed abundance trend is the 
large star-to-star scatter in 
[Ba/Eu] ratios. Whether this scatter is
real or due to inadequate Ba abundance analyses, however, is not clear.
Barium, with a number of isotopes and significant 
hyperfine and isotopic splitting,  can be difficult to analyze. 
Thus, for example, estimates of the proportions of the
relative isotopic abundances synthesized 
in the $r$- and $s$-processes are necessary for the abundance determination
(e.g. Magain 1995, Sneden et~al. 1996).
In addition, 
abundances of Ba  are also quite sensitive to the assumed 
values of the microturbulent
velocity.
These difficulties lead to typical uncertainties of at least 
$\sim\pm$0.2~dex 
in the [Ba/Eu] ratios in most metal-poor
stars, and thus make  this ratio   
inadequate to map out the detailed Galactic chemical 
evolution of the  
$r$- and $s$-processes (see Sneden et al. 2001a, for 
further discussion). 

As an alternative, La is also predominantly an $s$-process element in
solar system material (Burris et al. 2000).
Furthermore, with only one stable  isotope, 
La has more favorable atomic properties
than Ba does, making the abundance analysis more straight-forward. 
Thus, La/Eu could be utilized for chemical evolution studies.
For example, Johnson \& Bolte (2001) argue  that the  [La/Eu]
ratios indicate $r$-process dominance in Galactic stars as metal-rich as
[Fe/H]~$\sim$ --1.5.
A new large-sample study of La/Eu ratios in metal-poor
stars as a function of metallicity is being conducted 
(Simmerer et al., in preparation).
This survey  employs
higher resolution, higher S/N spectra and  the
new atomic data of Lawler {\it et al.} (2001).
Preliminary analysis 
indicates an excellent line-by-line abundance agreement 
for \ion{La}{2}
in the Sun, CS~22892-052, and BD+17~3248 (Sneden {\it et al.} 2001b).
In Figure 8 we show the behavior of [La/Eu] with [Fe/H] for
a small sample of  metal-poor but $n$-capture-rich stars.
We note that, unlike the case for the [Ba/Eu] ratios illustrated in 
Figure 7, there is a very small star-to-star scatter in [La/Eu] at 
lowest metallicities. However, the sample is still small, but
when the survey is completed 
it should be possible to ascertain with more certainty
the metallicity at which the $s$-process begin to contribute  
significantly to most stars' $n$-capture abundances.
Ultimately, such information should also help to determine
how rapidly  the Galaxy may have increased its Fe-peak
metallicity before the initial onset of the $s$-process from 
the deaths of the first intermediate-mass stars.


\section{NUCLEOCOSMOCHRONOLOGY WITH R-PROCESS CHRONOMETERS} 

The detections of thorium ($^{232}$Th) in several halo field stars 
and globular cluster stars make possible the use of its abundance, 
relative to stable $r$-process nuclei (e.g. Eu), to other long lived 
and spectroscopically observable 
radioactive species (e.g. $^{238}$U), or to its progeny ($^{208}$Pb) 
as a chronometer. 
Until recently the main emphasis has been on the use of 
the ratio Th/Eu, taking advantage of the fact that Eu is almost entirely 
produced in $r$-process nucleosynthesis. 
A promising new development has been the very recent
detection 
of both of the long-lived nuclear chronometers Th and U 
in one ultra-metal-poor 
(those with [Fe/H]~$\lesssim$ --2.5, hereafter UMP) halo star
(see Cayrel et~al. 2001).
These 
stellar chronometric- (or radioactive-) age estimates are critically
dependent upon accurate stellar abundance determinations and 
well-determined theoretical nucleosynthesis predictions of the 
initial abundances of the radioactive elements.
Currently the uncertainties in the age estimates employing this
technique are relatively large. It is expected, however, 
that these uncertainties will diminish with increasing numbers of 
stellar detections of 
Th and/or U and
of the stable \third 
$r$-process peak
elements - these latter measurements are needed 
to constrain theoretical abundance predictions of the radioactive chronometers.
Ultimately, the availability of improved determinations of 
the abundance of Pb (and perhaps Bi) will make possible even tighter 
constraints on stellar ages. In the following, we briefly review the current 
status of  these possible chronometers.

\subsection{Ages from Th/Eu}

The number of Th (produced solely in the
$r$-process) detections in metal-poor and 
UMP halo stars has been increasing
(Sneden et al. 1996; Cowan et~al. 1997;
Cowan et~al. 1999;
Sneden et al. 2000a; Westin et~al.
2000; Johnson \& Bolte 2001; Cayrel et al. 2001; 
Cowan et al. 2002).
These detections have led to 
direct chronometric
stellar age determinations. The typical technique has been to first 
determine the ratio of the abundance of 
radioactive thorium to the stable europium
abundance value.
Eu is a particularly good choice both because it is virtually a pure 
$r$-process nucleosynthesis product and because it 
is easily identified in the spectra of most 
metal-poor giant stars. These ratios are then directly compared with 
predictions of the  initial (zero-decay) value of
Th/Eu at the time of the formation of these elements. We illustrate this 
comparison in Figure 9, where the observed neutron-capture abundances 
in the UMP star CS~22892--052 (Sneden et al. 2000a) 
is compared with a scaled solar system
$r$-process abundance distribution (solid line). It is clear from this
comparison that the Th abundance has decayed (dropped) with respect to 
the stable heavy n-capture elements.  
Also shown in Figure 9 is a theoretical $r$-process abundance curve 
that reproduces the observed stable $r$-process elements and at the same
time (i.e., in the same calculation) predicts the initial radioactive 
elemental abundances.
Such predictions for initial abundance ratios of Th/Eu, or Th to 
other stable
elements, have been necessary since
the nuclei involved in the $r$-process are very far from stability,
and there are little experimental nuclear data available.
The difference in the initial (at time of formation) 
predicted ratio for Th/Eu and
the observed stellar ratio 
directly reflects (properly) 
the age of the 
radioactive 
material that was produced in the $r$-process site, whatever that may be.
However, 
since the time scale for formation of the current halo stars, after the 
ejection of this $r$-process material into the ISM,  would have been
relatively short (i.e., only millions of years, as compared to billions of 
years),  this age can be thought of as the stellar age. 
Chronometric ages, based upon the Th/Eu ratios, have 
typically fallen in the range
of 11--15 Gyr for the observed stars
(Sneden et al. 1996; 
Cowan et~al. 1997;
Pfeiffer, Kratz, \& Thielemann 1997; Cowan et~al. 1999;
Sneden et al. 2000a; Westin et~al.
2000; Johnson \& Bolte 2001; Cayrel et al. 2001;
Cowan et al. 2002). 

We also note that in one case Th/Eu ratios have been determined for
several giants in the globular cluster M 15
(Sneden et~al. 2000b), who estimated their average, and hence the
cluster,  age at 14 $\pm$ 4 Gyr.

Concern has been expressed, however, over the reliability of
employing Th/Eu as an age indicator in all metal-poor stars
(Goriely \& Arnould 2001).
In particular,  
Eu is widely separated in mass number from thorium and 
differences in the synthesis histories of these elements could lead
to age uncertainties. Furthermore, the initial predicted values of
this abundance ratio  are very sensitively
dependent upon the choice of nuclear mass formula -- very different
mass formulae lead to widely differing initial ratios and hence
age estimates. 
Particularly troubling is the case of CS~31082--001. This UMP star
shows very high values of Th and U with respect to the stable elements
such as Eu (Cayrel et al. 2001; Hill et al. 2002). Schatz et al. (2002)
in fact show that employing Th/Eu in this star leads to an unrealistically
young age, while Th/U provides an answer (15.5 $\pm$ 3.2 Gyr) that is consistent
with other age estimates for these metal-poor stars. 
It is not clear, at this point, whether CS 31082--001 is
somehow unusual or whether is represents a different class of stars. 
The results for this star do suggest though caution in 
employing only Th/Eu for chronometric stellar age estimates.
Clearly,  obtaining abundances of stable elements nearer in mass number
to thorium, or even better, obtaining two chronometers such as 
Th and U would be preferred for this radioactive dating technique.

\subsection{Ages from Two Chronometers: Th/U}

The recent, and first, detection of uranium in the UMP star
CS~31082-001 ([Fe/H] = --2.9)  by Cayrel et al. (2001) offers  promise
for these stellar age detections with the addition of a second chronometer.
An added advantage is that 
uranium is nearby to thorium in nuclear mass, and thus
might offer a more reliable comparison than lower mass (i.e., farther
away) elements such as, for example,  
europium. 
Cayrel et al. (2001) (see also
Schatz et al. 2002)
employed the abundances of U and Th, in combination
with each other and with some other stable elements,
to find an average chronometric age of 12.5  $\pm$ 3 Gyr
for CS~31082-001. 

Recent observations of the spectrum of BD +17$^o$3248 
also indicate the presence of uranium, albeit weakly,
marking this as the second metal-poor halo star in which U has been
detected (Cowan et al. 2002).
These authors, employing Th, U, Eu and several \third $r$-process
peak elements, estimate the age of  BD +17$^o$3248 to be 13.8 $\pm$ 4 Gyr. 
All of these age calculations make use of 
the theoretically predicted initial (zero-decay) values for ratios
such as Th/U, Th/Eu and Th/Pt. These in turn are based upon 
nuclear mass formulae and abundance fits to the heaviest stable
neutron-capture elements, i.e., the \third $r$-process peak elements. 
Several nuclear mass formulae,
such as the extended Thomas Fermi
model with quenched shell effects far from stability, i.e.,
ETFSI-Q (see Pfeiffer et al. 1997 for discussion),
have been employed in making the 
age determinations. 
(See also Schatz et al. 2002 for new age determinations of CS~31082-001
employing another nuclear mass model.)
The number of abundance determinations of the \third $r$-process peak 
elements, including Pt, in metal-poor stars is also increasing, particularly
as the result of the utilization of the Hubble Space Telescope (HST) --
the dominant transitions of elements such as Pt are primarily in the 
ultraviolet requiring a space-based  telescope. Such accurate stellar 
elemental abundances 
are  important in ratio to the observed radioactive abundances and  
in constraining the predicted initial (time-zero)
radioactive abundances employed in these  chronometric-age determinations.

We caution, however, that even with these recent observational successes,
uranium detections in metal-poor or UMP stars 
may  turn out to be uncommon.  Previous 
observations have failed to detect U in such 
stars as CS~22892--052 and HD~115444, although meaningful upper limits
can also be employed to constrain the age estimates
(see also Burles et al. 2002 for further discussion). 
Finally, the overlap among the various chronometric
age estimates for the field halo stars and for the globular M15
with other globular cluster (e.g., 14 Gyr from Pont et al. 1998)
and cosmological age determinations
(e.g., 14.9 $\pm$ 1.5 Gyr from Perlmutter et al. 1999 and 14.2 $\pm$ 1.7 Gyr
from Riess et al. 1998) is encouraging.
While the chronometric age predictions have suffered from some uncertainties
in the past, 
that situation has been slowly changing with the addition of more 
high resolution stellar  data and  
more nuclear
data leading to increasingly more reliable
prescriptions for very
neutron-rich  nuclei
(see Pfeiffer et al. 2001; Burles et al. 2002).


\section{NEUTRON-CAPTURE NUCLEOSYNTHESIS AT VERY LOW METALLICITIES} 

In this section we examine the differences in  abundances
among $r$-process ``rich'' and ``poor'' stars as a further probe of the 
early history of  Galactic nucleosynthesis.  
(We define $r$-process rich as [Eu/Fe] $>$ +0.7 and poor as [Ba/Fe] $<$ --0.7.) 
We compare abundances of these two groups for selected elements in 
Tables 1 and 2. A range
in metallicity  is seen in both groups with most of these stars having
[Fe/H] between  about --2.0 and --3.1,
and that $r$-process ``richness'' does not 
correlate directly with metallicity --
the $r$-process poor stars are not necessarily the most metal-poor.

The behavior of the Sr/Ba ratio, as a function of 
[Ba/Fe], is also illustrated in Figure 10 to compare and contrast
these groups of stars.  The $r$-process rich stars in Figure 10 generally
have [Ba/Fe] $>$ 0, or supersolar.
The Sr/Ba ratios for these 
stars cluster around the pure $r$-process ratio.
The situation is quite different for the $r$-process poor stars. 
They all have [Ba/Fe] ratios much below the solar value,
typically 1/10 $^{th}$ or less. On the other hand, those same stars
have very large values of [Sr/Ba] with ratios 10--100 times solar,
in contrast to the subsolar values for the $r$-process rich stars.
While the weak $s$-process in massive stars (discussed previously) 
can augment the synthesis of Sr,
the key finding here is the extremely low values of the heavier 
neutron-capture elements, e.g.,  Ba, leading to the very large ratios of 
Sr/Ba observed in $r$-process poor stars.

This comparison of $r$-process-rich and $r$-process-poor stars
suggests that the nucleosynthetic production (in an $r$-process 
environment) of the lighter elements, such as  
Sr, and probably Y and Zr,  is  favored over the 
the heavier elements such as Ba in the progenitors of some metal-poor stars.
Production of  the heavier 
$r$-process elements 
early in the history of the Galaxy may be delayed relative to the lighter
ones. 
In this scenario 
the observed $r$-process poor  stars may be 
showing the products of nucleosynthesis
from progenitor stars that lived and died prior to the formation of the 
first ``main'' $r$-process stars. If lower mass (8--10 \Msun) supernovae 
are a likely source for the main $r$-process (see e.g.,
Mathews et al.   1992; Wheeler, Cowan, \& Hillebrandt
1998; Ishimaru \& Wanajo 1999),
then the high Sr/Ba
ratios could be  reflective of higher mass star ejecta.  
Both the lighter and heavier neutron-capture elements might have been
synthesized in some type of incomplete or abbreviated $r$-process in
early, extremely metal-poor massive stars, 
with a possible additional Sr synthesis from the
weak $s$-process in these same stars. 
Such an incomplete $r$-process, with insufficient neutron flux to flow 
up to the heaviest neutron-capture elements can occur   in 
some model calculations for certain astrophysical conditions 
(see  Burles et al. 2002 for further discussion).
It is also possible that some other element besides iron, 
for example silicon, 
could act as a seed for any  initial $r$-processing 
(see Hannawald, Pfeiffer, \& Kratz 2001).
Thus, even in very metal-poor or UMP stars (with by definition
low iron abundances) some $r$-processing could occur that might synthesize
some levels of Sr and even Ba.
Heavy element synthesis has even been predicted for certain types of 
inhomogeneous big bang models (see Rauscher et al. 1994). 
Some of these models, under certain very restrictive assumptions, 
can produce $s$-like, as opposed to $r$-like, abundance signatures for 
neutron-capture elements that would favor lighter neutron-capture
element production.

Detailed abundance determinations for large numbers of neutron-capture
elements in individual $r$-process poor stars have, in general, not 
been available. The low levels of the abundances and the weak lines
have made such element detections difficult. One star that has been
extensively studied is the bright giant HD 122563. In Figure 11 we show 
an abundance comparison between this star and four $r$-process rich
stars ( BD +17$^o$3248, CS 22892--052, CS 31082-001, and HD 115444). 
For each star we show the difference between the abundance value of
the element and the corresponding solar system $r$-process prediction.
A perfect agreement would result in a relatively flat-line curve. That is
in general, what is observed for the four $r$-process rich stars,
supporting earlier arguments
that these neutron-capture elements in these stars are in scaled solar system 
$r$-process proportions.  
The abundance data, what there is of it, for HD 122563 does not follow
the same pattern, however. Instead the abundances of the heavier neutron-capture
elements seem to depend on, and decline with,  atomic number, Z. 
This may,  in fact,  be another indication of an incomplete $r$-process
synthesis from a massive star early in the history of the Galaxy. 
We caution, however, that our analysis is based  so far only  
on  HD 122562, and needs to be confirmed by additional stars.

While the most detailed abundance data is so far only available for HD 122563,
there have been other recent papers noting somewhat similar abundance traits 
in some of the other very metal-poor stars.  In particular,  a few of these
early Galactic stars show high concentrations of C, N or even O along with
very low levels of heavier neutron capture elements (see e.g., 
Norris, Ryan \& Beers 1997, 2001;  Depagne et al. 2002). 

In sum these brief comparisons suggest that
the early phases of Galactic nucleosynthesis are likely to be complex.
As evolutionary time scales become  much shorter than 
Galactic mixing time scales, the yields from different (progenitor) 
mass-range stars may show up as different chemical mixes, so that 
the interpretation of abundances in stars that are more metal-poor than, 
for example, [Fe/H] = --3 may be problematic in the context of chemical
evolution. Additional detailed abundance determinations, particularly 
for the more metal-poor  and $r$-process poor stars, may help
to clarify these issues in the future.

\section{CONCLUSIONS}


The detailed nature of the processes by which the heavy elements from iron  
to uranium are made in nature stands as one of the most fundamental and 
challenging unsolved problems in nuclear astrophysics. While the  
$s$-process is understood to occur primarily 
in the helium burning 
shells of AGB stars, the site of the $r$-process can only be 
constrained by theory 
to a rapidly evolving stellar/supernova 
environment, sufficiently short lived to explain the presence of 
$r$-process nuclei in the very oldest stars studied in our Galaxy. 
Furthermore, neither the nuclear physics of the $r$-process of 
neutron-capture synthesis nor the physical properties of the environment in 
which it might operate are known to sufficient reliability to allow us to 
understand the extraordinary robustness of the $r$-process 
abundance pattern in the mass region A $\gtaprx$ 130-140.

Enter the observational astronomers. Over the past two decades, high resolution 
spectroscopic studies have provided high quality data that has helped chemical 
evolution and nucleosynthesis theorists by imposing constraints upon both the 
nature of the nucleosynthesis sites and the abundance history of the early 
Galaxy. On the basis of the observed trends we have reviewed in this paper, 
we can summarize the following conclusions:

\begin{itemize}

\item[]{1.} Observations leave absolutely no doubt but that the first 
contributions to the abundances of the heavy $r$-process elements 
in our Galaxy occurred at an early epoch, and well in advance of the first 
substantial $s$-process contributions. At metallicities 
[Fe/H] $\ltaprx$ --2.5, the elements in the range  Z $\ge$ 56 
 are virtually pure 
$r$-process products. $r$-Process synthesis of 
A $\gtaprx$ 130-140 isotopes happens early in Galactic history, prior to 
contribution of heavy $s$-process isotopes  from AGB stars.

\item[]{2.} The $r$-process mechanism for the synthesis of the 
A $\gtaprx$ 130-140 isotopes (the ``main'' or ``strong'' component) 
is extremely robust. This 
is reflected in the fact that the abundance patterns in the most
metal-deficient (oldest) stars, which may have received contributions from 
only one or at best a few $r$-process events, are nevertheless 
indistinguishable from the $r$-process abundance pattern that 
characterizes solar system matter. (The abundance patterns for the three stars 
CS 22892-052, HD 115444, and BD +173248 shown in Figure 2 reveal this 
remarkable agreement for stars of low metallicity but high 
[$r$-process/Fe].) 

\item[]{3.} The Ba/Eu and particularly La/Eu ratios reveal that the first 
significant (or major) 
introduction of heavy $s$-process isotopes (the Ba peak 
nuclei and beyond: the main $s$-process component) 
occurred at a metallicity [Fe/H] $\sim$ -2. The time 
at which this occurred, presumably set by the lifetime of the low mass 
($\sim$ 1-2 M$_\odot$) AGB star $s$-process nucleosynthesis site, is 
of order 10$^9$ years.

\item[]{4.} The increased scatter in [$r$-process/Fe] at low 
metallicities [Fe/H] presumably reflects a significant and increasing level 
of inhomogeneity present in the gas at that epoch. It also provides  
evidence for the fact that only a fraction of the massive star 
(M $\gtaprx$ 10 M$_\odot$) and associated Type II supernova environments can 
have contributed to the synthesis of the heavy $r$-process isotopes. 
The data shown in Figure 4 reveal levels of $r$-process 
enrichment relative to iron of factors of approximately 50.

\item[]{5.} The observations, particularly in $r$-process rich-stars, 
indicate that the
heavier (Ba and above, Z $\ge$ 56, or A $\gtaprx$ 130-140) 
neutron-capture elements are 
consistent with a scaled solar system $r$-process curve. The data, 
although still incomplete, seem to indicate that the lighter neutron-capture
elements (for example Ag) are not consistent with (i.e., fall below) that
same scaled $r$-process curve. This behavior is shown in Figure 2 for 
CS~22892-052,  and   
indicates that two  distinct $r$-process environments may 
be required to synthesize both ends of the abundance distribution. 
These observations support earlier suggestions of two $r$-processes based
upon solar system meteoritic (isotopic) data (Wasserburg et al. 1996). 
Analogously to the $s$-process, 
we can attribute the heavier neutron capture elements to a ``strong''
robust $r$-process, with a ``weak'' $r$-process responsible for the
synthesis of the lighter elements below barium.
Analyzing the lighter abundance data 
is complicated by the possibility of the production of 
$s$-process nuclei (in the weak $s$-process) from massive stars 
that might contribute to the production
of Sr, Y and Zr only.  
Additional  spectroscopic studies of the abundances of the 
lighter neutron-capture elements,  
in metal deficient stars, will be  required to sort all of this out. 

\item[]{6.} The identification of the $r$-process site with massive 
star environments implies that the critical nuclear chronometers for dating 
the Galaxy were formed early in Galactic history. This strongly supports the 
use of the $r$-process isotopes $^{232}$Th, $^{235}$U, and $^{238}$U 
as reliable chronometers of the Galactic nucleosynthesis era. A more detailed 
discussion of this and related issues will be provided in a forthcoming 
paper (Burles {\it et al.} 2002). 

\item[]{7.} For the $r$-process poor stars at low metallicity the data seem
to indicate that lighter elements such as  Sr  
have high abundances with respect to
heavier neutron-capture elements such as Ba. 
Also, for these stars (at least based
upon the data available for the star HD 122563), the abundances of the 
heavier neutron-capture elements seems to depend more on atomic number
(and to decline faster with atomic number) than the standard $r$-process
production normally does. These results suggest that the early nucleosynthesis
history of the Galaxy was quite complex with yields
coming from various $s$- and $r$-process components and from synthesis 
sites with a variety of 
progenitor mass ranges.

\end{itemize}

\acknowledgments

We thank Jennifer Simmerer for providing us with her La/Eu data 
in advance of publication and
we thank Faith Jordan for assistance with the Sr/Ba analysis.
This research has been supported in part by 
STScI grants GO-8111 and GO-08342,
NSF grants
AST-9986974  (JJC),
AST-9987162  (CS),
and by the ASCI/Alliances Center for Astrophysical Thermonuclear Flashes
under DOE contract B341495 (JWT).
C. Pilachowski gratefully acknowledges the support of Indiana University
through the Research Fund of the Daniel Kirkwood Chair in Astronomy.

\clearpage
\pagestyle{empty}

\tablecolumns{6}
\tablewidth{0pt}

\begin{deluxetable}{lrrrrrrr}
\tablecaption{Neutron-Capture Abundances in r-process Poor Stars}
\tablehead{
\colhead{Star}                          &
\colhead{[Fe/H]}                        &
\colhead{[Sr/Fe]}                       &
\colhead{[Ba/Fe]}                       &
\colhead{[Sr/Ba]}                       &
\colhead{[Eu/Fe]}                       &                          
\colhead{[La/Fe]}                       &                       
\colhead{[La/Eu]}                       \\ 
}
\startdata
HD 4306        &   --2.54 & --0.20  & --1.25 &   1.05   & \nodata  & \nodata   & 
\nodata\\
HD 88609       &   --2.93 &   0.30  & --0.91 &   1.21   & \nodata  & \nodata   & 
\nodata\\
HD 122563      &   --2.71 &   0.17  & --0.93 &   1.10   &  --0.30  &  --0.71   &  
--0.41\\
BD -18 5550     &   --2.93 &   0.18  & --1.02 &   1.20   &  --0.07  & \nodata   & 
\nodata\\
CS22897-008  &   --3.35 &   0.63  & --1.23 &   1.86   & \nodata  & \nodata   & 
\nodata\\
\enddata

\end{deluxetable}

\clearpage
\pagestyle{empty}

\tablenum{2}
\tablecolumns{6}
\tablewidth{0pt}

\begin{deluxetable}{lrrrrrrr}
\tablecaption{Neutron-Capture Abundances in r-Process Rich Stars}
\tablehead{
\colhead{Star}                          &
\colhead{[Fe/H]}                        &
\colhead{[Sr/Fe]}                       &
\colhead{[Ba/Fe]}                       &
\colhead{[Sr/Ba]}                       &
\colhead{[Eu/Fe]}                       &                          
\colhead{[La/Fe]}                       &                       
\colhead{[La/Eu]}                       \\ 
}
\startdata
HD 20      &   --1.39 &   0.05  & 0.58   & --0.53   & 0.77  & 0.39   & --0.38\\
HD 126587      &   --2.85 & --0.45  & --0.06 & --0.39   & 0.75  & 0.30   & 
--0.45\\
BD +17 3248     &   --2.1  &   0.09  & 0.40  & --0.21   & 0.91  & 0.45   & 
--0.46\\
CS22892-052   &   --3.1  &   0.70  & 0.41   & --0.26   & 1.70  & 1.07   & --0.63\\
HD 115444      &   --3.0  &   0.32  & 0.18   &   0.14   & 0.85  & 0.37   & 
--0.48\\
CS31082-001   &   --2.9  &   0.65  & 1.17   & --0.52   & 1.63  & 1.13   & --0.50\\
\enddata

\end{deluxetable}

\clearpage
\pagestyle{empty}


\begin{figure}
\plotone{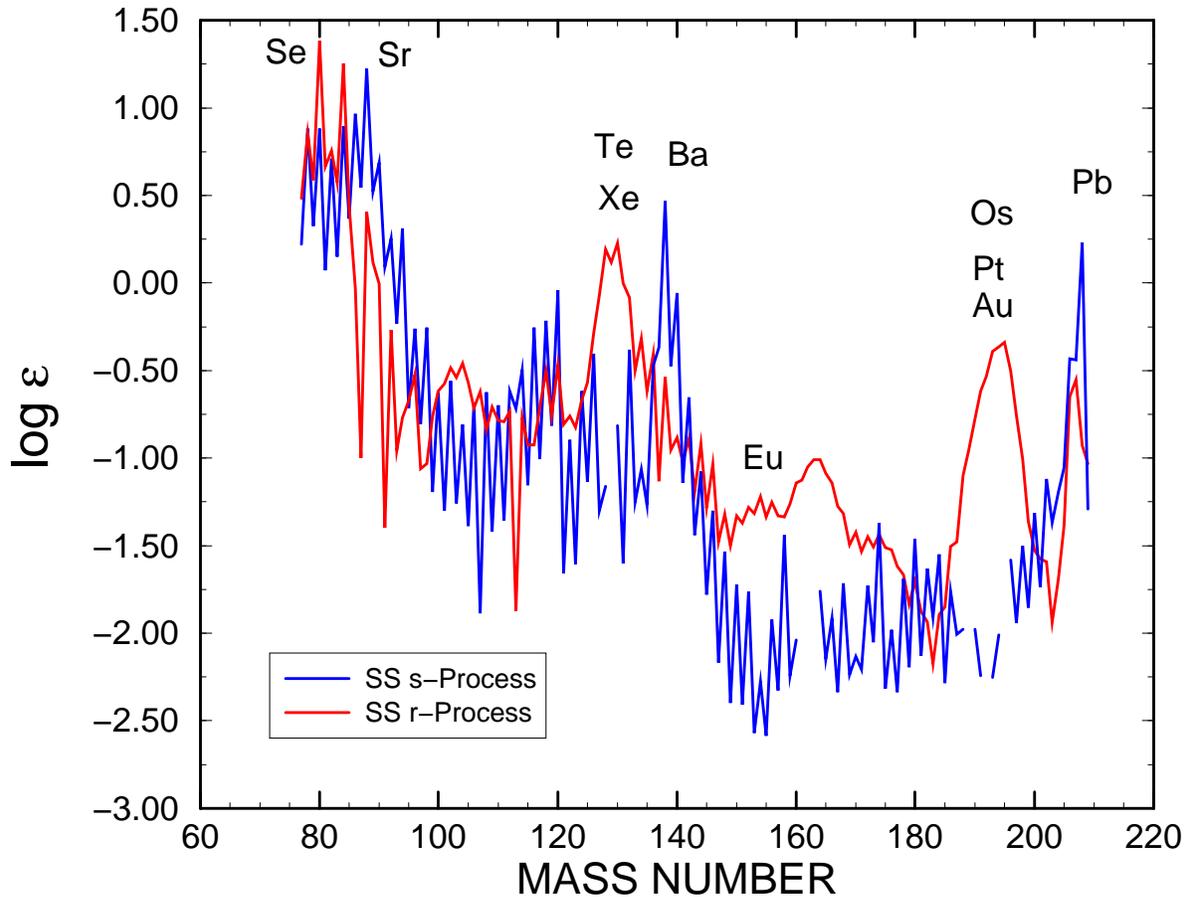}
\caption{
The $s$-process and $r$-process
abundances in solar system matter,
based upon the work by K\"appeler {\it et al.} (1989). Note the
distinctive $s$-process signatures
at masses A $\approx$ 88, 138, and 208 and
the corresponding
$r$-process signatures at A $\approx$ 130 and 195, all
attributable to closed shell effects on neutron capture cross sections. It is
the $r$-process pattern thus extracted from solar system abundances
that can be compared with the observed heavy element patterns in extremely
metal-deficient stars. The total solar system abundances for the heavy 
elements are those compiled by Anders and Grevesse (1989). 
\label{fig1}}
\end{figure}

\clearpage
\begin{figure}
\epsscale{1.0}
\plotone{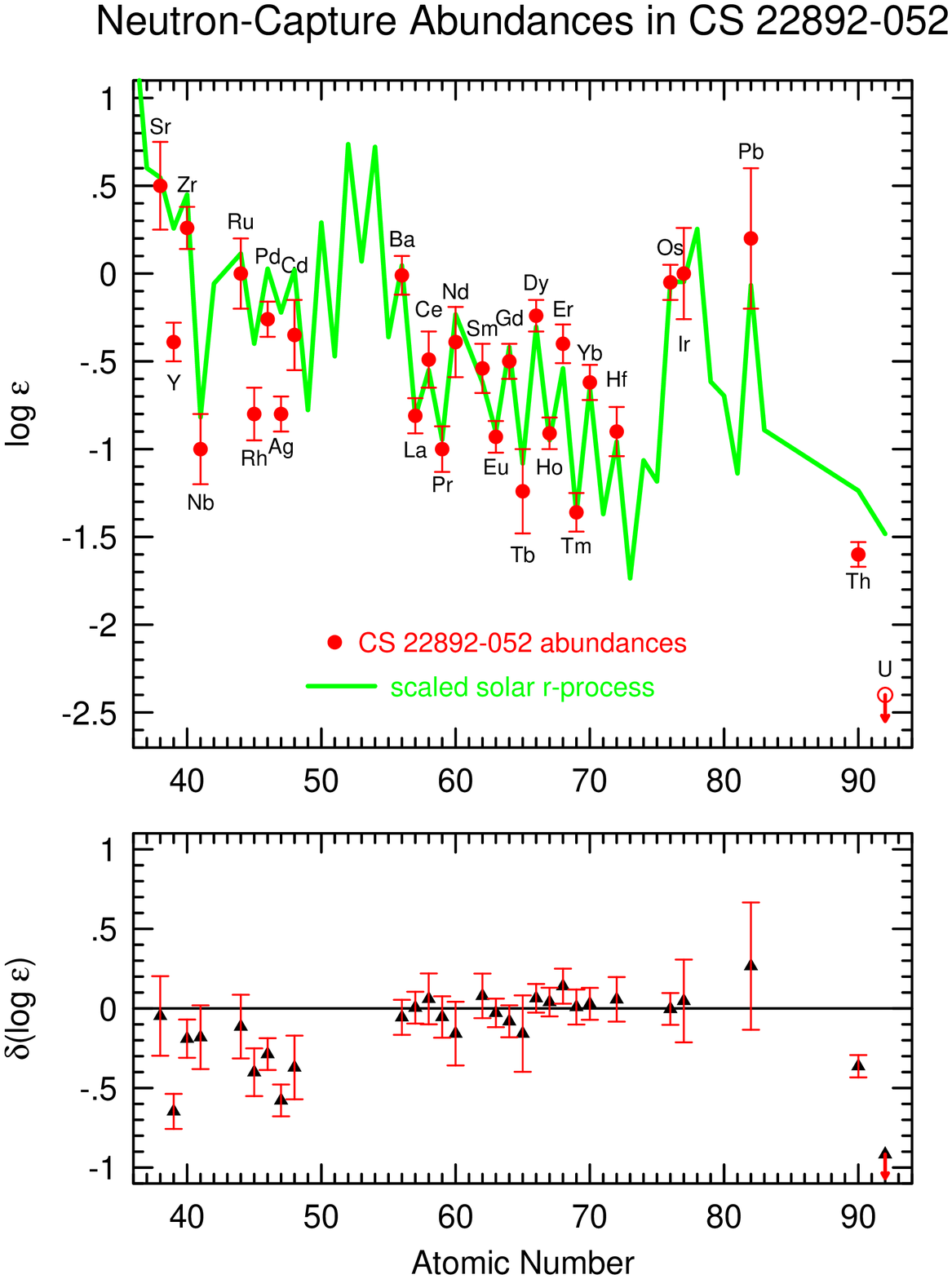}
\vskip -0.4truein
\caption{The total heavy element abundance patterns for CS 22892-052 is
compared with the scaled solar system
$r$-process abundance distribution
(solid line). (Sneden {\it et al.} 2000a)
\label{fig2}}
\end{figure}

\clearpage
\begin{figure}
\plotone{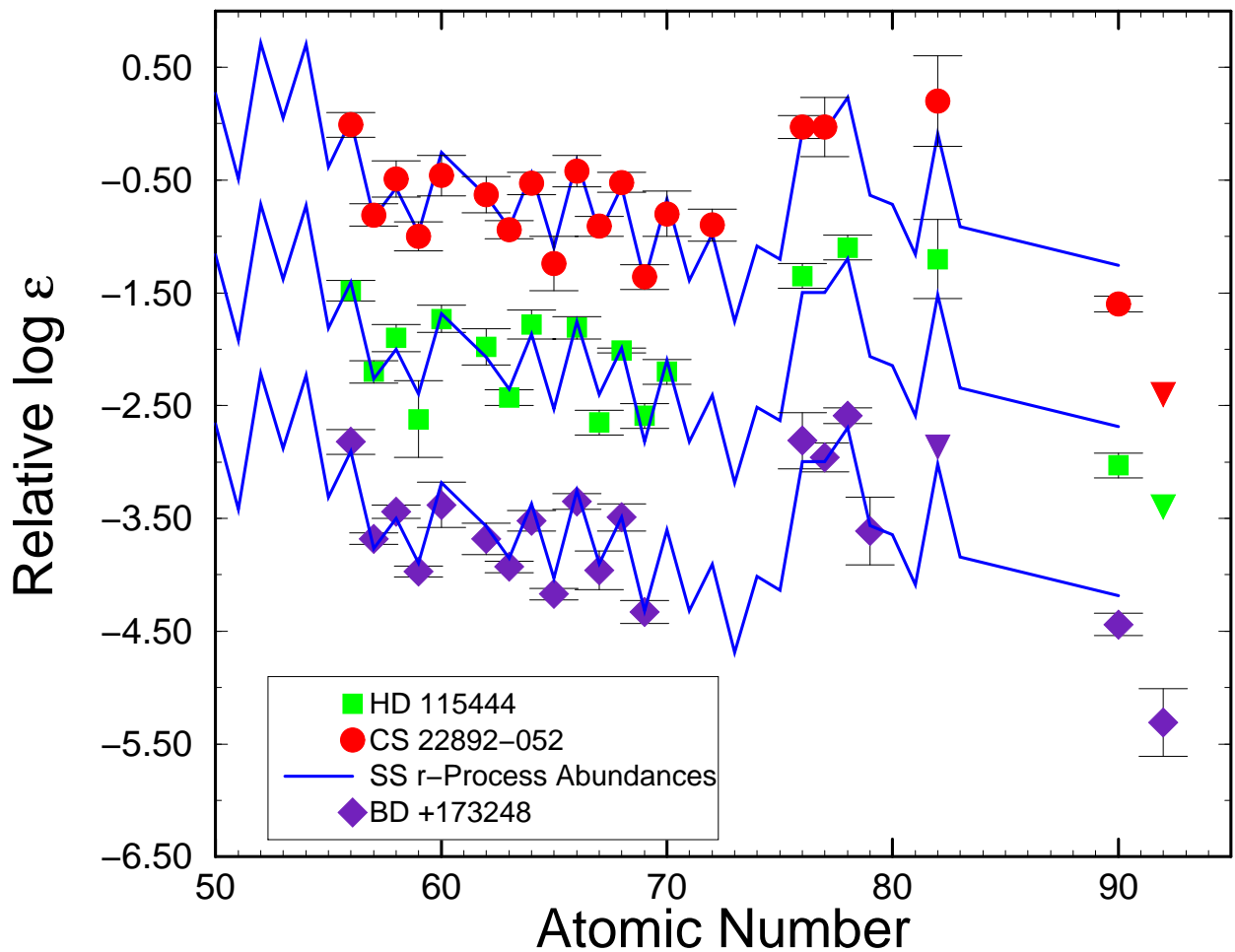}
\caption{
The heavy element abundance patterns for the three stars 
CS 22892-052, HD 155444 and BD +17$^o$3248 are compared 
with the scaled solar system 
$r$-process abundance distribution (solid line). (see Sneden {\it et al.} 2000; 
Westin {\it et al.} 2000; Cowan et al. 2002) Upper limits are indicated
by inverted triangles. 
\label{fig3}}
\end{figure}

\clearpage
\begin{figure}
\plotone{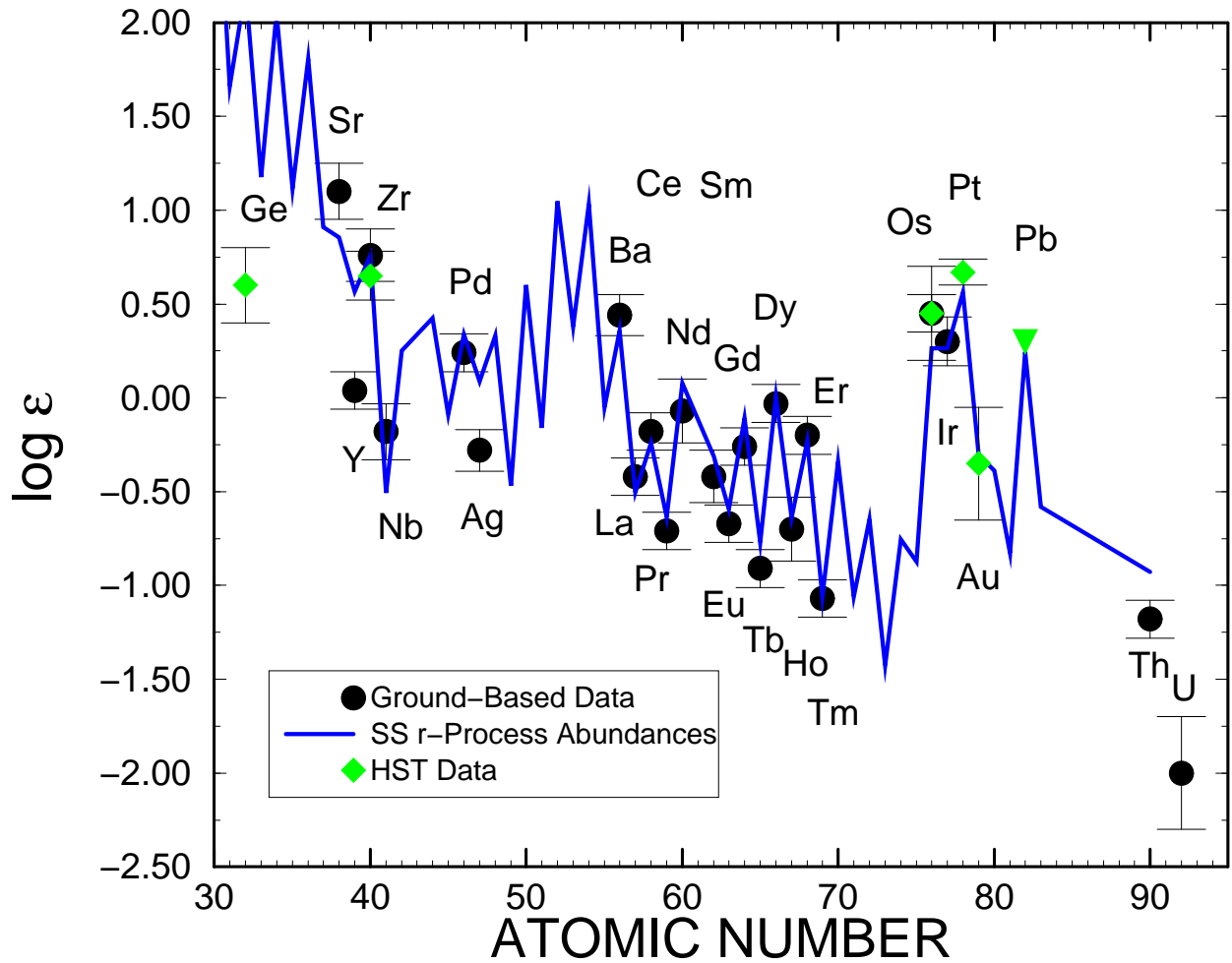}
\caption{
Neutron capture elements in the halo star BD +17$^o$3248 (Cowan {\it et al.} 
2002), obtained from ground based and HST observations, are compared to a 
scaled solar system $r$-process abundance curve. The upper limit 
on the lead abundance is denoted by an inverted triangle. Note also the 
thorium and uranium detections.
\label{fig4}}
\end{figure}

\clearpage
\begin{figure}
\epsscale{0.9}
\plotone{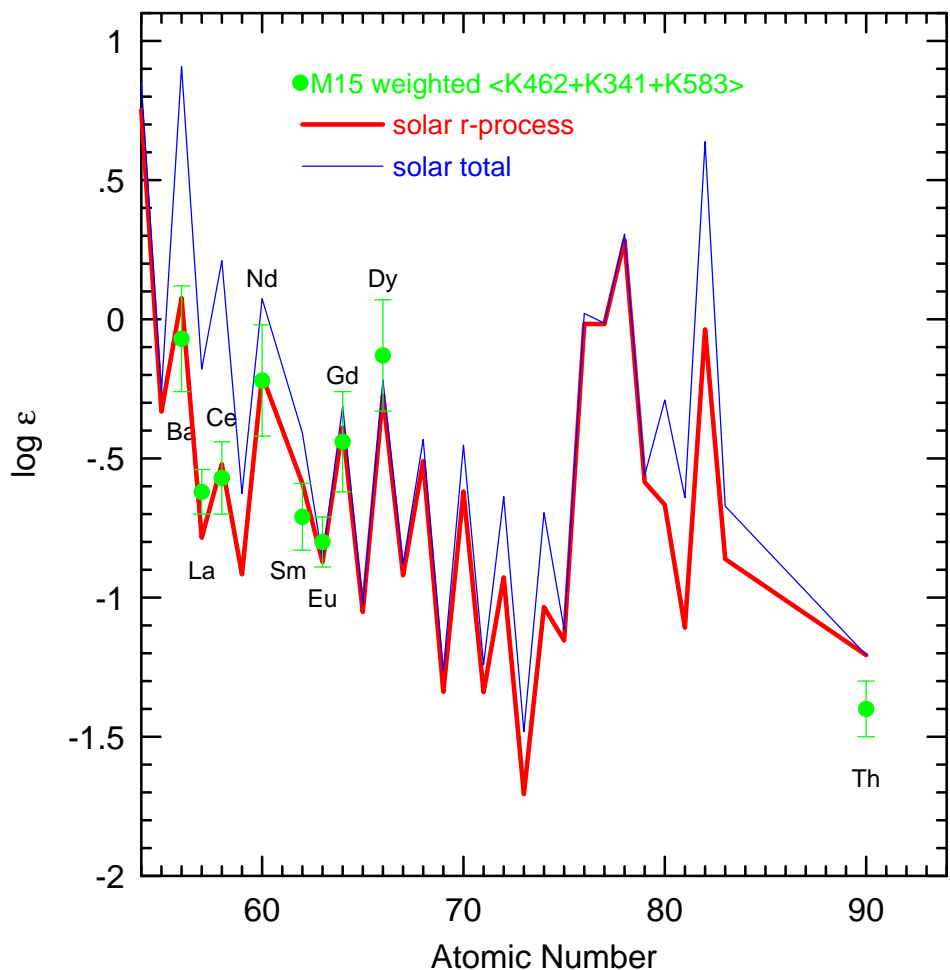}
\vskip -1.2truein
\caption{
Weighted mean abundances of n-capture elements with
Z~$\geq$~56 in M15 giants.
In forming the means, the abundances of K462 are given triple weight,
K341 double weight, and K583 single weight, to account for the
relative [n-capture/Fe] abundance levels of these three stars.
The solar-system $r$-process-only abundance distribution (Burris \etal\
2000) has been shifted by its mean difference compared
to the M15 points shown in this figure for elements Ba through Dy,
and plotted as a solid  curve.
The solar-system total abundance distribution has been shifted by
this same amount and plotted as a dotted  line.
(Sneden {\it et al.} 2000b)
\label{fig5}}
\end{figure}

\clearpage
\begin{figure}
\plotone{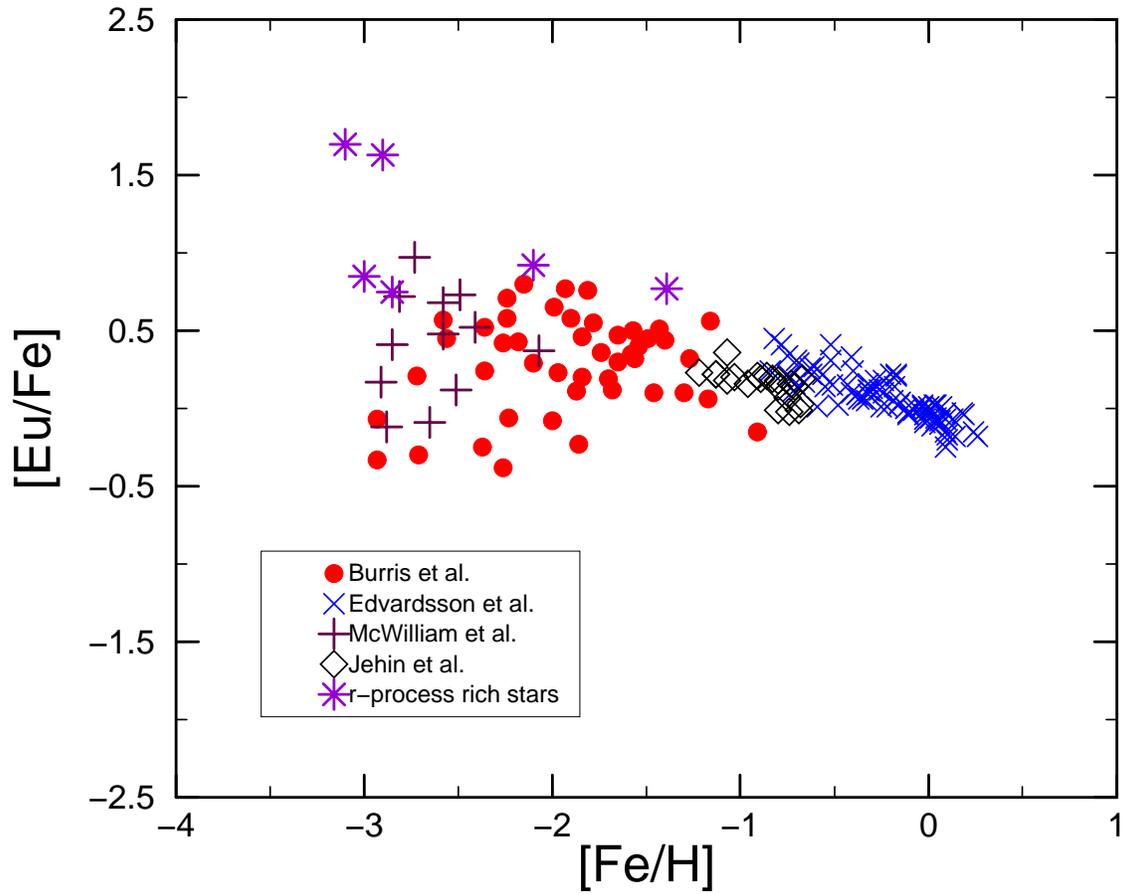}
\caption{
The ratio [Eu/Fe] is displayed
as a function of [Fe/H]
for a large sample of halo and disk stars (Burris {\it et al.} 2000;
Jehin et al. 1999; Edvardsson et al. 1993; McWilliam et al. 1995, 
McWilliam 1998). 
$r$-process rich stars from Westin et al. (2000);
Sneden et al. (2000a), Hill et al. 2002 and Cowan et al. (2002).
\label{fig6}}
\end{figure}

\clearpage
\begin{figure}
\epsscale{1.0}
\plotone{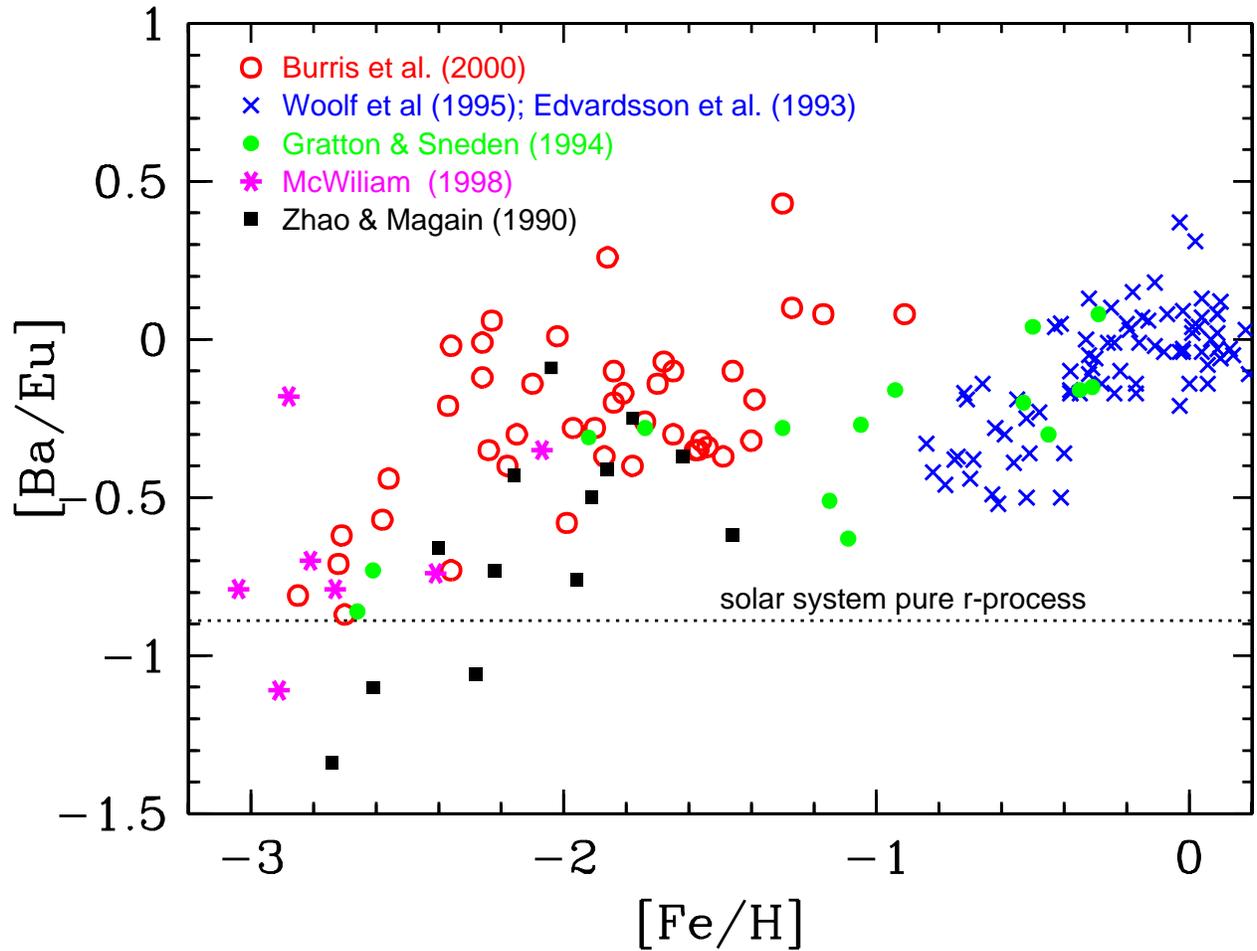}
\vskip 0.2truein
\caption{
The ratio [Ba/Eu], which reflects the ratio of
$s$-process to
$r$-process elemental abundances,
is displayed as a function of [Fe/H] for
a sample of halo and disk stars. 
\label{fig7}}
\end{figure}

\clearpage
\begin{figure}
\epsscale{1.0}
\plotone{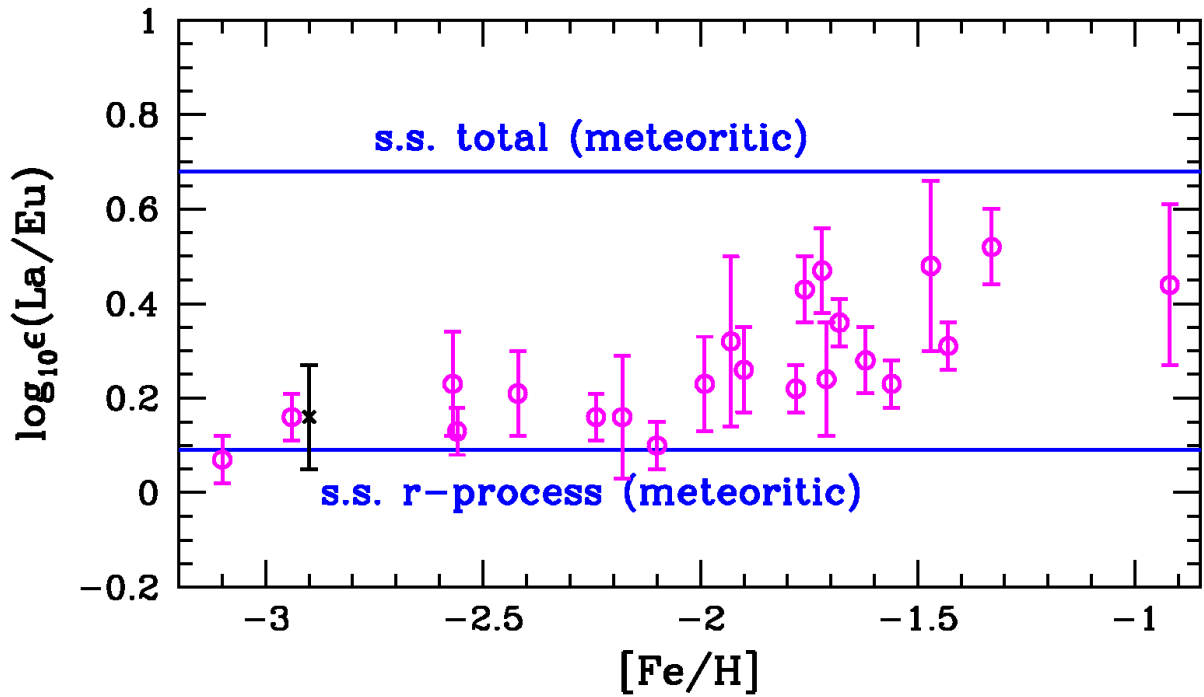}
\vskip -.3truein
\caption{
Ratio of La to Eu abundances in a few representative stars (open circles) over
most of the Galactic halo metallicity range from Simmerer et al. (2002). 
The La abundances have been
derived using new La II laboratory values from
(Lawler  {\it et al.} 2001).
The data point indicated by an x is
from Cayrel et al. (2001).
\label{fig8}}
\end{figure}

\clearpage
\begin{figure}
\plotone{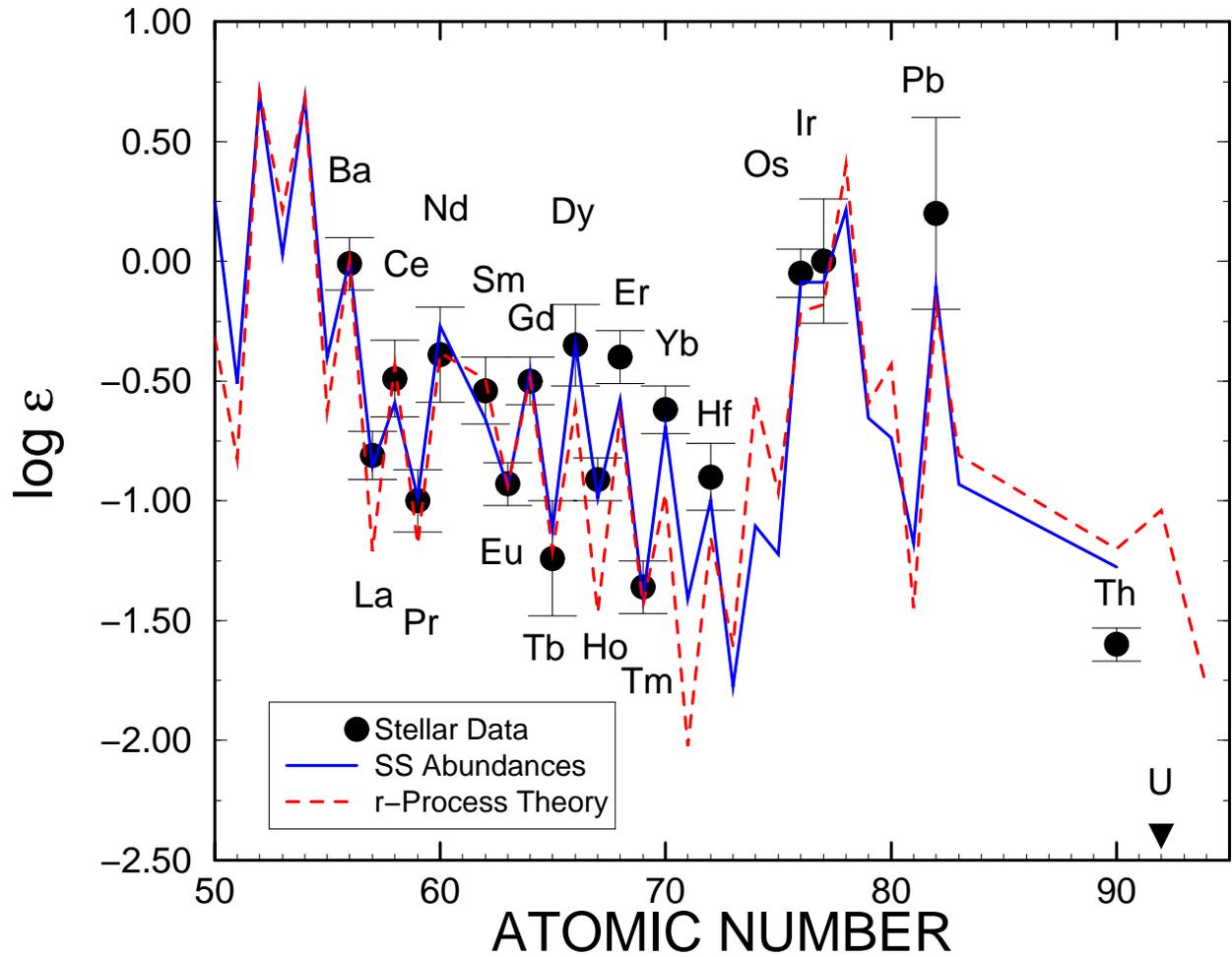}
\caption{
The heavy element abundance patterns for CS 22892-052 is
compared with the scaled solar system
$r$-process abundance distribution
(solid line). (Sneden {\it et al.} 2000). Also shown is a theoretical $r$-process
abundance curve (dashed line) from Cowan et al. (1999).
\label{fig9}}
\end{figure}

\clearpage
\begin{figure}
\plotone{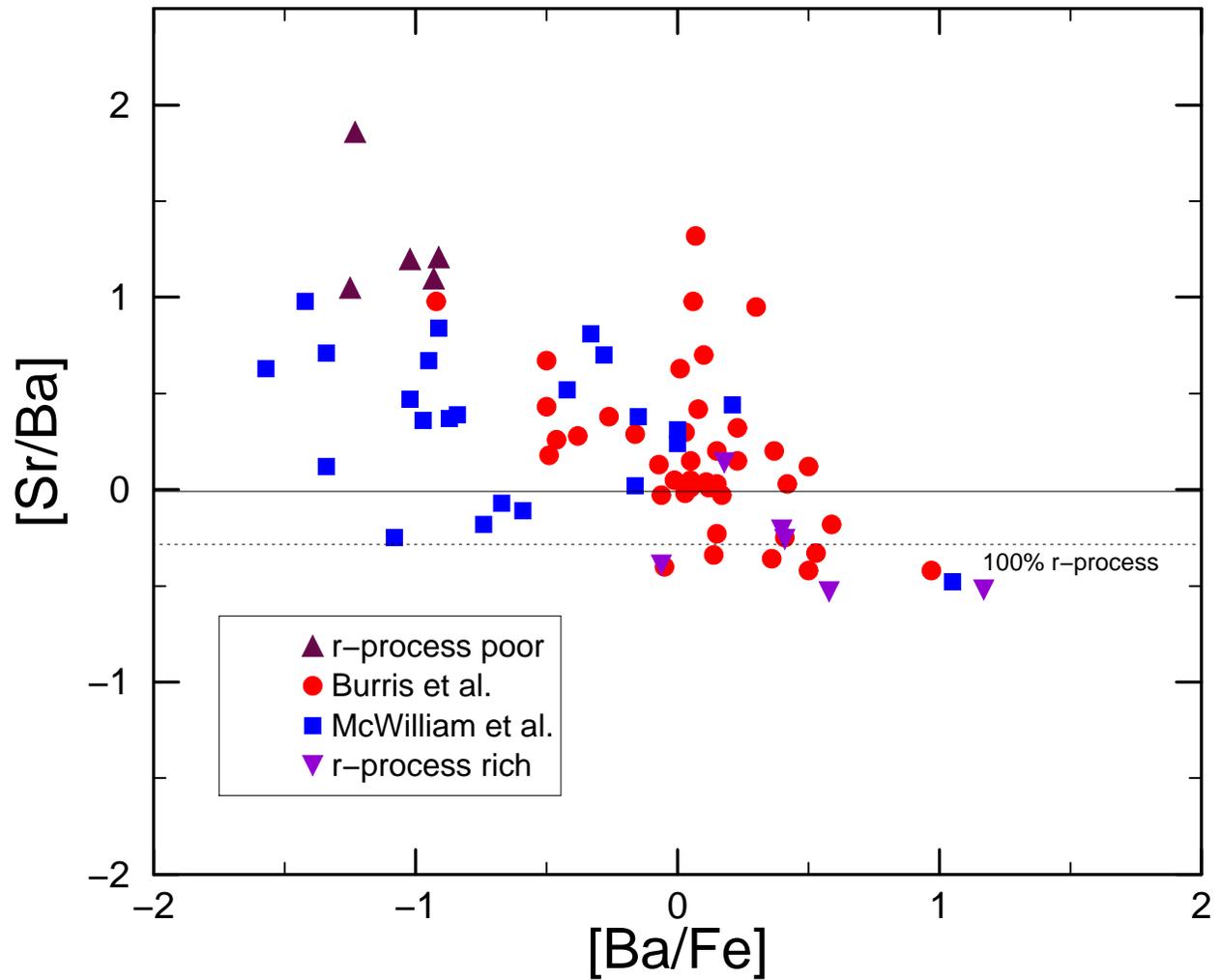}
\caption{
The ratio [Sr/Ba] 
is displayed as a function of [Fe/H] for
a sample of halo and disk stars 
(Burris {\it et al.} 2000;
Jehin et al. 1999; Edvardsson et al. 1993; McWilliam et al. 1995,
McWilliam 1998).
Note in particular the $r$-process poor and $r$-process rich stars.
\label{fig10}}
\end{figure}

\clearpage
\begin{figure}
\epsscale{0.8}
\plotone{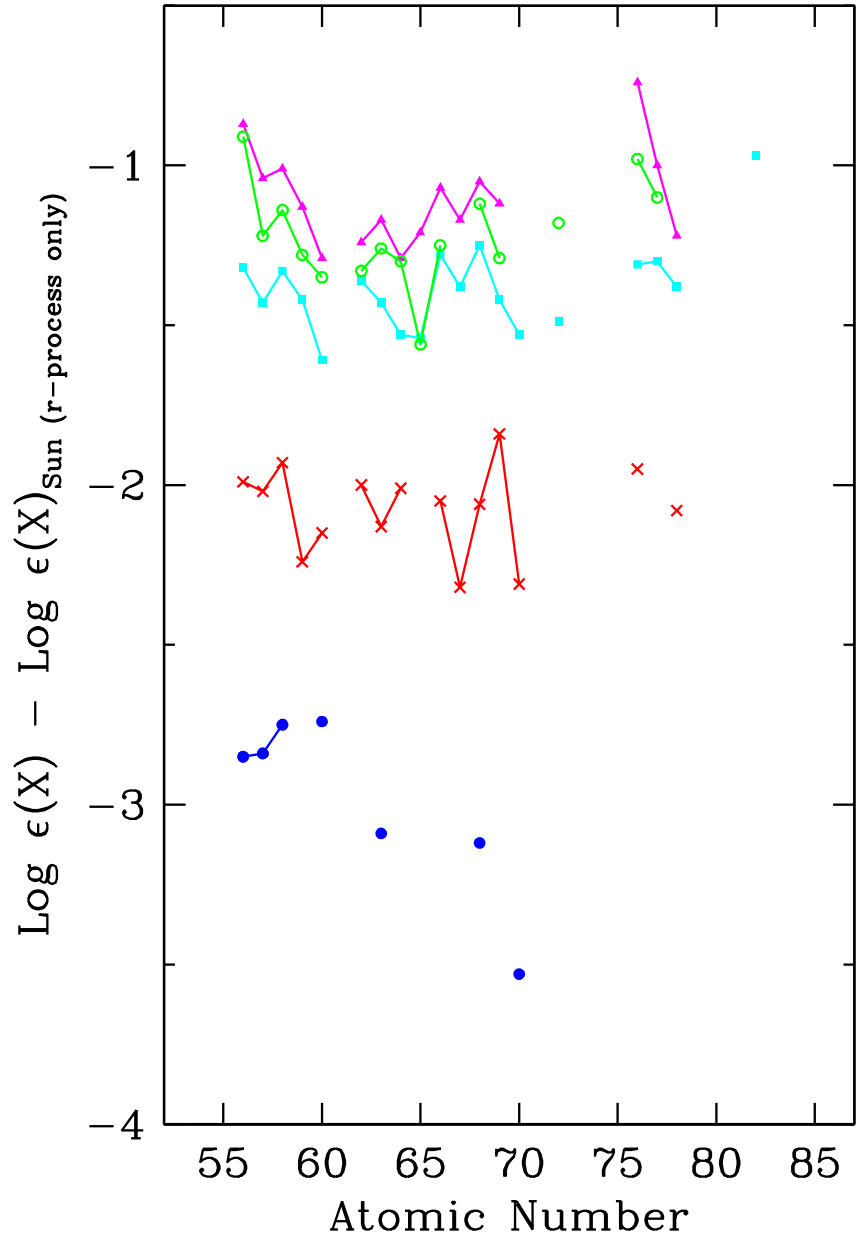}
\caption{
Comparisons of the abundance patterns between four  $r$-process rich 
(bd+17 3248, triangles,
CS 22892-052, squares,
HD 115444, crosses, 
CS 31082-001, open circles)
and the
$r$-process poor star HD 122563
(filled circles).
\label{fig11}}
\end{figure}

\end{document}